\documentclass[preprint]{aastex}
\usepackage{graphics}

\slugcomment{To appear in the Astronomical Journal}

\shorttitle{Brown Dwarf Variability} \shortauthors{Enoch et al.}

\begin{document}


\title{Photometric Variability at the L/T Dwarf Boundary}


\author{Melissa L. Enoch} 
\affil{Department of
Astronomy, California Institute of Technology, MS 105-24, Pasadena, CA
91125} \email{menoch@astro.caltech.edu}

\author{Michael E. Brown} \affil{Division of Geological and Planetary
Sciences, California Institute of Technology, MS 150-21, Pasadena, CA
91125} \email{mbrown@caltech.edu}

\and

\author{Adam J. Burgasser\altaffilmark{1}} \affil{Department of
Astronomy and Astrophysics, University of California, Los Angeles,
CA, 90095-1562} \email{adam@astro.ucla.edu} 

\altaffiltext{1}{Hubble Fellow}




\begin{abstract}
We have monitored the photometric variability of nine field L and T brown 
dwarfs for 10 nights during the course of one month.
Observations were obtained in the $K_s$-band with the Palomar 60 inch 
Telescope Near-Infrared Camera.
Results of statistical analyses indicate that at least three of the nine 
targets show significant evidence for variability, and three more are 
possibly variable.  Fractional deviations  from
the median flux vary from 5\% to 25\%.   Two of the variable targets, 2MASS
0030-14 (L7) and SDSS 0151+12 (T1), have marginally  significant peaks
in their periodograms.  The phased light curves show evidence for
periodic  behavior on timescales of 1.5 hours and 3.0 hours respectively.   
No significant correlations between variability amplitude and spectral 
type or $J-K_s$ color are found.  While it is clear that variability exists 
in objects near the L/T dwarf boundary, we find no evidence that variability 
near the L/T boundary is more likely than it is for early L dwarfs. 
\end{abstract}




\section{Introduction}\label{intro}

In the last several years two new low mass stellar and brown dwarf
spectral classes, L and T dwarfs, have been  established (see,
e.g. \cite{mart97, mart99, kirk99, geb02, burg02b}).  Now that a
method exists by which to classify these objects, efforts have turned to
understanding their structure, origin,  and evolution.   One means of
probing the atmospheric structure of L and T dwarfs has focused on the
presence, or  lack thereof, of photometric variability, which may be tied to 
magnetic activity or cloud properties \citep{ack01}.

Grain formation is likely an
important process  in the atmospheres of cool stellar and sub-stellar
objects, providing a significant source of opacity
(e.g. \cite{lun89}).  Iron and magnesium silicates such as enstatite
should condense out and form droplets below about $2200$K, leading to
the formation of clouds \citep{burr99}.  Current atmospheric
calculations (e.g. \cite{ack01, all01}) find that early L
dwarf spectra  are best matched by a model atmosphere with thin
silicate and iron clouds, late L dwarfs by an atmosphere with thick 
clouds, and T dwarfs
by a clear atmosphere in which clouds presumably have fallen below
the region from which flux emerges.  The change in $J-K_s$ colors from
progressively redder for later L types  \citep{kirk99, kirk00, leg01,
leg02} to relatively blue for early T types \citep{leg99, leg02,
burg02b}  also suggests a substantial atmospheric change across the L/T
boundary.  By assigning effective temperatures to each spectral
subtype, \cite{kirk00} predict that the  transition from L to T takes
place over a very small range in effective temperature ($\le350$K).

\cite{burg02a} have attempted to describe the transition in cloud cover 
that appears to occur
at the  L/T transition by appropriately interpolating between clear and 
cloudy models \citep{ack01}.
The authors find that such partly cloudy atmospheres can
reproduce the colors and magnitudes of late L and  early T dwarfs, and
suggest that substantial variability around $\lambda=1\mu$m might be 
observed in this region due to the breakup of cloud layers, the appearance 
of long-lived cloud holes, or a combination of these effects.  

It is unclear on what timescale such variability would
occur, or how large a fractional  variability amplitude to expect.  A single
large, long lived feature or cloud hole might cause variability  with
a stable period, rapidly changing features would cause
variability with no  discernible period, and uniformly distributed
features would not produce a variability  signal at all \citep{bjm01,
gel02}.
As noted by \cite{coop02}, theoretical modeling of such a transition would 
require 3D cloud models, including convection,  which do not yet exist for 
objects beyond the solar system.

Previous work has found several M and L stars and brown dwarfs to be
photometrically variable  in the optical \citep{tinn99, bjm99, bjm01,
mart01, clark02a, clark02b, gel02}, but  no dwarfs later than L5 were
included in these samples.  \cite{nak00}  report NIR
spectroscopic variability in the T6 dwarf SDSS 1624+00,
while \cite{kirk00} find spectroscopic variability in the L8 dwarf Gl 584C.

The observed variability could be a consequence of cloud 
structure, magnetic
activity or spots,  or some unknown phenomenon.  Most
variability studies, however, favor the cloud interpretation  
\citep{mart01,tinn99, gel02}.
\cite{gel01} argue that magnetic spots are probably not responsible
for photometric  variability in L dwarfs due to the low magnetic
Reynolds number in their cool atmospheres.   Given that detections of
H$\alpha$, one indicator of magnetic activity, drops dramatically in
later L and T dwarfs \citep{kirk00, giz00}, (There are, however, a few 
notable exceptions -- \cite{burg02c}),  it seems unlikely that magnetic spots 
could cause variability in very late L or T dwarfs.

In an effort to better understand variability around the L/T boundary,
we have  undertaken a program to photometrically monitor one peculiar early L,
four late L, and four early T field dwarfs.  In Section~\ref{obssec} we describe 
the observations and reduction of data.  We present light curves for all nine 
targets and the results of variability analyses in Section~\ref{res}.  In 
Section~\ref{discussion} we briefly discuss our results in the context of ideas 
about cloud-breakup at the L/T boundary, and Section~\ref{sum} gives a short summary.

\section{Observations \& Data reduction}\label{obssec}

Targets were observed during 2001 September 30-October 6 and October
25-28 (UT), providing a total observational baseline of 29 days for each 
target.  The first
observation  period, September 30-October 6, will hereafter be referred
to as Set~1, October 25-28 as Set~2,  and the complete duration of
observations as the combined set.   Observations were made with the
$K_s$ ($\lambda=2.17\mu$m) filter of the Near-Infrared Camera  at the
Palomar Observatory 60 inch Telescope.  The instrument is a cassegrain
256$\times$256 pixel camera with a pixel scale of 0.619 arcsec/pixel
and total field of view of 6.98 square  arcmin.  Bright
late L and early T dwarf targets were chosen
from the 2 Micron All Sky Survey (2MASS, \cite{skrut97}), the
Sloan Digital  Sky Survey (SDSS, \cite{york00}), and the Deep Near
Infrared Survey of the Southern Sky (DENIS, \cite{epch97}).
All of the targets have assigned spectral types between L6 and T5 with
the exception of 2MASS 2208+29, which is an L2p.\footnote{This brown
dwarf is classified as peculiar because it has TiO bands similar to an
L4 dwarf, but KI lines similar to an L0 dwarf \citep{kirk00}.  Note that 
we assume that all objects later than L6 are substellar; as the earlier 
objects have lithium detections \citep{kirk00}, we refer to all the targets 
as brown dwarfs.}
Properties of the targets appear in Table~\ref{tbl1}.

Observations were made in the $K_s$ band only, as many of the L dwarfs
were too faint  to be imaged in the $J$ band with this telescope.
The targets were observed in a cyclic manner, each being observed a
total of 2-5 times per night.   During each observation, 18 images of
the target were taken on a $3\times3$ square dither  pattern (2 images
at each position).\footnote{Due to the positioning of the one
reference star in  the field for 2MASS 0328+23, this object was
observed on a $2\times2$ square dither pattern with 4  images at each
position.}  All exposures were 30 seconds in duration.  Weather was
mostly clear with some patchy clouds on several of the nights; October
2  was cloudy and yielded no usable data.  Conditions were not always
photometric, but for the following analysis only relative magnitudes between 
objects within an image were computed.  Therefore, only if  the average flux changed
significantly during an 18 image set, or if clouds were too thick as
determined by eye, were any image sets rejected.  The average seeing for
all nights was $1\arcsec-1.\arcsec8$ (2-3 pixels).

The raw images were first dark subtracted and flat fielded.
Dome flats, twilight flats, and dark frames were
obtained each night.   Flat frames for each night were constructed by
a median combination of the dome  flats; as there appeared to be no
residual structure in the images, the twilight flats  were not used.
Bad pixel masks were created to flag outliers and dead pixels in  the
flat field.  Images were subsequently sky subtracted using a median
combination of the dithered  science images, and reduced images
were  then aligned using relative positions of the brightest objects in
the field.

Aperture photometry was computed for all objects in each field using 
a 5 pixel aperture, with the sky value  computed in a circular annulus.  
Due to the small field of view, there was generally only one bright star 
available in each field.  Therefore, differential
photometry ($\Delta$m) of targets was calculated with reference to a single 
bright star in the field (hereafter the comparison star).  
Differential photometry for all other stars in the field (hereafter field stars) 
was calculated with reference to this same comparison star.
For two of the targets only one other star (which is used as the comparison star) 
lies within the frame, so no field star is available.  For the majority of  
other targets, field stars are comparable in brightness to the target.

\section{Results}\label{res}

We assume that no variation is taking place on the 10 minute
timescale of each  set of 18 dithered images.  A relative magnitude
difference ($\delta$m) is calculated  between the target and
comparison star for each of the 18 images, by the method described in Section~\ref{obssec}.  
The mean of these, $\Delta$m = $\overline{\delta\mathrm{m}}$, is computed at each
observation  time interval, and the light curve constructed.  One
sigma error bars represent the  error in the mean as computed from the
scatter in $\delta$m ($\sigma_{\Delta\mathrm{m}}^2=\sigma_{\delta\mathrm{m}}^2/N$).
Thus the error bars on $\Delta$m are computed empirically from the data 
themselves, and no estimation of systematic errors has been made.  This 
technique has the advantage that false detections associated with underestimating 
the errors present (excluding systematics) can be avoided.   
However, variations on short timescales (such that 
there is substantial change in the light curve in 10 minutes) will result 
in artificially large error bars, masking any variability signal present.  In 
some cases not all 18 images are included in $\Delta$m because patchy clouds 
or changes in the seeing made accurate photometry of the faint targets 
impossible.

\begin{figure}
\plotone{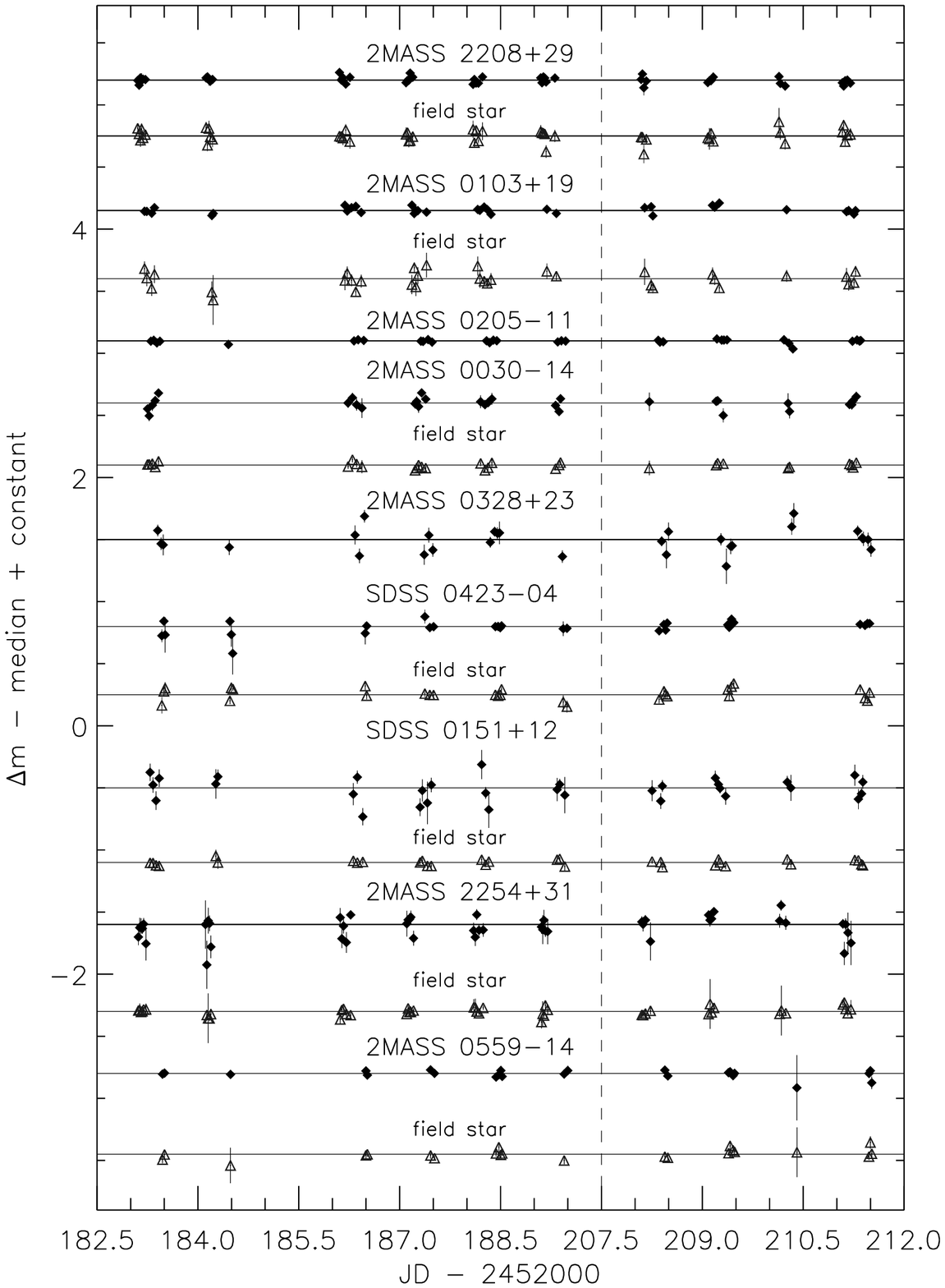}
\caption{Relative photometry for all targets and field stars.
Target data appear as solid diamonds and field star data as open 
triangles.  For some targets no field star is available.  
Note the break at JD=2452207.5 (indicated by the dashed line), 
where 420 hours between the two data 
sets have been removed for clarity. \label{lightall}}
\end{figure}

Light curves based on relative photometry for each target are shown in
Figure~\ref{lightall},  where the median value of
$\Delta$m has been subtracted from the data and an arbitrary
constant added.  Shown below each target is the light curve of a field star 
(computed using the same comparison star as for the target), when available.  
Note that a 420-hour time  span (from JD=2452190.0 to JD=2452207.5) between 
Set~1 and Set~2 has been cut from the  x-axis for clarity.

\subsection{Variability}

To determine if any of the targets show evidence for variability we
define a robust median statistic:
\begin{equation}
\eta = \sum_{i=1}^N \left|
\frac{\Delta\mathrm{m}_i-median(\Delta\mathrm{m}) } { \sigma_i }
\right|.
\end{equation}
We choose to use a robust method to evaluate variability rather than
the  $\chi^2$ test because robust methods are much less sensitive to
any non-Gaussian errors  that may be present\footnote{As a consistency 
check we have also done a $\chi^2$ analysis following \cite{bjm99, bjm01}, with very 
similar results to our robust method.}.  
We define the reduced
$\eta$ to be $\widetilde{\eta} = \eta/d$ where d is the number of degrees
of freedom.  If the data are not varying, i.e. the intrinsic
underlying  distribution is a constant value, then the expected value
of $\widetilde{\eta}$ is  less than one. A value of 
$\widetilde{\eta}\gtrsim 1$ indicates that an object is likely to be
variable.  To quantify this likelihood we ran Monte Carlo simulations.

The significance of a given $\eta$ is  determined by
replacing the data with random noise (with a mean of zero  and a
standard deviation of one) and computing the resulting $\eta_r$.  This
process  is repeated 1000 times; the number of times that these random
$\eta_r$ are smaller than the real $\eta$ determines the
confidence level.  For example, if  980 of the random $\eta_r$ are
smaller than $\eta$, then $\eta$ has a confidence of
98\%.  We take a confidence above 95\% to indicate significant
evidence for variability.

The results of the robust analysis -- the values of $\widetilde{\eta}$
and the confidence (conf) for each of the targets -- are presented in 
Table~\ref{tblvar}.
For comparison, variability results for the field stars are also shown.  
Note that if a comparison star were responsible for the variability seen 
in a target light curve, we would expect the field star to also show variability.
The variability analysis is done for  Set~1 and Set~2 separately, as
well as for the combined set.  Detections, i.e. those objects  with
conf~$>95\%$, are shown in bold-faced type.  Detections of which we are less 
confident for some reason (see Section~\ref{possvar}) are labeled possible 
detections and left in plain type.

The light curves were also examined to determine if there were any
statistically  significant changes in target brightness between
Set~1 and Set~2.  The two sets were compared using the Student's T
statistic and distribution function.  None of the targets
showed a statistical difference between the data sets with a confidence 
above 95\%.

In order to compare the results for different objects it is necessary to 
 know the minimum amplitude at which variations could have been detected 
in each light curve.  We choose detection limit of 99\% because 
we want a stringent restriction on possible variations in the field star 
light curves, and thus comparison star variability.  If variability were 
present at an amplitude above the 
detection limit it almost certainly would be detected.  
Sinusoidal detection limits are determined by generating sinusoidal
data with a given amplitude  and a period of 4.0 hours, which is close
to the rotational period expected for L dwarfs from  rotational
velocity measurements \citep{bas00}.  Changing the period chosen for this
analysis does not greatly affect the detection limits, although  in
general shorter periods yield slightly lower limits.  The signal is sampled 
at time intervals matching the 
actual observations, and random errors with standard
deviations equal to the true errors in the data are added.  This
process is repeated 100  times at each amplitude and the fraction of
times that the signal produces a confidence $>95\%$ is  recorded.  The
amplitude at which 99 of the 100 test light curves are flagged  as a
detection is the 99\% detection limit.   

Detection limits for random variations are determined in a similar
way.  Rather than a sinusoid, a series of random numbers with zero mean 
are generated, for which the amplitude is given by the largest
peak-to-peak spread.  While we would not actually expect our targets to vary 
in a random manner, we consider randomly varying light curves as an extreme 
case of non-sinusoidal variability.  We note that it is generally easier to detect 
a sinusoidal variation than a random one, and  the detection limits in
Set~2 are usually higher than for Set~1, due to the smaller number of
data points  in Set~2.

The 99\% detection limits and approximate amplitudes of variability, where
applicable, for targets and field stars are listed in
Table~\ref{tblamps}.  For detections and possible detections, approximate
peak-to-peak amplitudes are given in magnitudes, with
uncertainties.  Peak-to-peak amplitudes range from
0.10 to 0.48 mags,  corresponding to approximate fractional
deviations from the mean flux of 5\% to 25\%.  The uncertainty in the 
peak-to-peak amplitude is determined by adding in quadrature the 
$1\sigma$ errors of the two data points used to determine the amplitude 
(the highest and lowest points in the light curve).
Detection limits for both sinusoidal and random variations are given as 
peak-to-peak amplitudes (again in mags)
for all targets and field stars.

\subsubsection{Discussion of individual objects: Detections}\label{detvar}

\textbf{2MASS 0030-14:}  A confidence of 98\% makes this L7 dwarf a
fairly strong detection in Set~1, but it is  only a marginal detection
(conf~$=95\%$) in the combined set (see Table~\ref{tblvar}).  
An examination of the light curve combined with the fact that the detection 
limit in Set~2 (0.22 mags) is greater than the  amplitude in Set~1 (0.19 mags), 
makes it apparent that Set~2 could have been a detection were more data 
available.  The 99\% detection limits  for the field star (0.07
mags for sinusoidal variation and 0.11 mags for random variations)
are about 1$\sigma$ below the detected amplitude of variation for
2MASS 0030-14 ($\sim0.19\pm0.11$ mags).   Furthermore, as we will see in
Section~\ref{persec}, the variations in 2MASS 0030-14 appear to be
close to sinusoidal.  As no such behavior is seen in the field star
light curve, this strengthens our confidence that the variability is
intrinsic to 2MASS 0030-14.

\textbf{SDSS 0151+12:}  This T1 dwarf is a detection in Set~1 (99\%) 
and marginally in the combined set (95\%).  The estimated amplitude  
($0.42\pm0.14$ mags) is well above the field star detection limits 
of 0.07 mags (sinusoidal variations) and 0.12 mags (random variations).  
Thus it is unlikely that the observed variability is due to the comparison
star.  No evidence for variability is seen in Set~2 (conf~$=64\%$), even 
though the detection limits in Set~2 (0.30 and 0.33 mags) are less than 
the Set~1 amplitudes (they are, however, within $1\sigma$).  This suggests
that the amplitude of variation may have decreased between the two sets.
As with 2MASS 0030-14, evidence for periodic variations in 
Section~\ref{persec} strengthens our confidence in this detection.

\textbf{2MASS 2254+31:}  The T5 dwarf 2MASS 2254+31 is a detection in
the combined set (98\%), primarily due to an increase in flux during the 
second and third nights of Set~2.  This two night trend is not seen in
the field star light curve or that of any of the other targets, although 
there are similar deviations in 2MASS 0328+23 and 2MASS 0205-11 on the 
third night only.  There is no detection in Set~1, despite the 
sinusoidal detection limit (0.21 mags) being more than
$1\sigma$ lower than the  estimated amplitude in Set~2 ($0.48\pm0.20$).  Given that 
the random variation limit (0.33) is less than $1\sigma$ below the amplitude, 
it seems likely that the variations are non-sinusoidal.

\subsubsection{Discussion of individual objects: Possible detections}\label{possvar}

\textbf{2MASS 0103+19:}  This L6 dwarf is detected as variable
in Set~2 and the combined set, due to a short rise in
flux at around JD=2452209.3.  When the highest point is removed 2MASS
0103+19  drops just below the detection limit (conf$=92.3\%$ in Set~2
and 88.8\% for the combined  set).  Detection limits for the field star 
($\sim0.2$ mags sinusoidal, 
$\sim0.3$ mags random) are substantially greater than the observed
peak-to-peak  amplitude of variability ($\sim0.1\pm0.02$ mags), so we
are unable to rule out the  possibility that the detected variability
may be due to the comparison star.  We note,  however, that if we use the
field star as the comparison star we still find variability in 2MASS 0103+19
with conf~$=95\%$ for Set~2.  The high detection limits and larger 
observed scatter in the field star light curve are due to the fact that 
the field star is almost one magnitude fainter than 2MASS 0103+19.   
Given the uncertainties present we label 2MASS 0103+19 a possible detection.

\textbf{2MASS 0328+23:}  2MASS 0328+23, an L8 dwarf, has one of the
largest amplitudes of variability for  the detections here
($\sim0.43\pm0.16$ mags for the combined set), and a very high confidence 
($>99\%$ in Set~1 and the combined set).  The light curve (Figure~\ref{lightall}) 
illustrates that the scatter in the data is much greater than the error bars.
As there is no field star available, however, it is impossible to tell
whether this variability is intrinsic to the brown dwarf or to the 
comparison star.

\textbf{SDSS 0423-04:}  SDSS 0423-04 is a detection in the combined
set with a confidence of 96\%.  It is clear, however, from the field star 
values in Table~\ref{tblvar} that either the comparison star
or field star is variable in this case (conf~$=97\%$ for Set~1 and
conf~$>99\%$ for Set~2 and the combined set).   
When we examine the aperture photometry (i.e. the apparent magnitude m, 
\textit{not} $\Delta$m) of the target, the field star and the comparison star, 
we find that the trends in the field star light curve that cause it to be 
flagged as variable seem to be caused by the field star.  
Thus we believe that it is likely the field star, not 
the comparison star, which is variable.  
If the field star is used as the comparison star, SDSS~0423 becomes a 
detection in Set~2 only.  It is still unclear whether or not SDSS 0423-04 
is actually the source of these variations, thus it is a possible detection.

\subsection{Periodicity}\label{persec}

When significant evidence for variability is found, we check for the
existence of periodic  variations by constructing the the Lomb-Scargle
weighted periodogram for unevenly sampled data  \citep{wood92}.  
Given the large gap between the data
sets, we choose to analyze the Set~1 and Set~2 data independently.   
In computing the periodogram we allow for a maximum period of 150 hours 
(approximately equal to the observational baseline for Set~1)
and a minimum period of 1.0 hour (approximately the Nyquist sampling limit).  
Peaks in the periodogram indicate
how well  a sinusoid with a given period fit the data.
The statistical significance of these peaks are determined in the
following way:  the  data are shuffled into random order, keeping
the same observation times.  This  shuffling is repeated 1000 times
and the highest periodogram peak for each recorded.  The percentage  
of shuffled peaks below a given value determines that value's significance.

\begin{figure}
\plotone{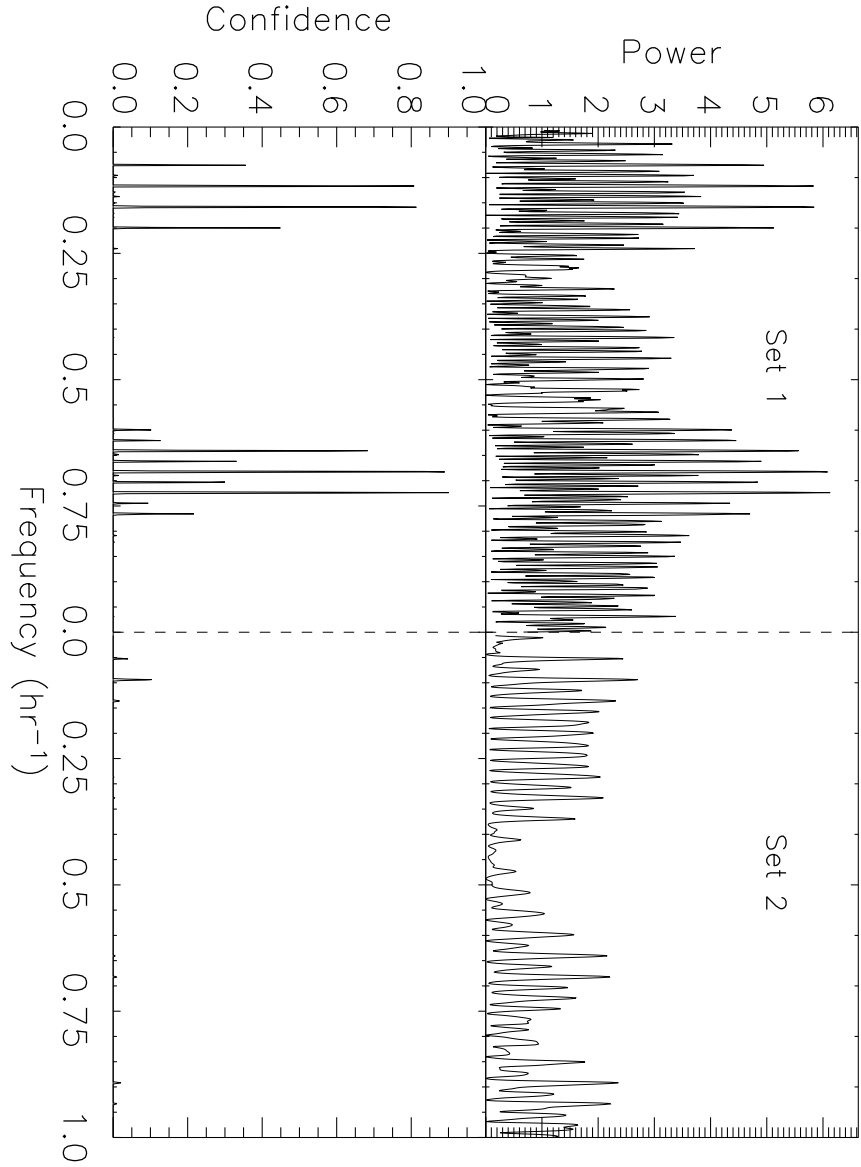}
\caption{Periodogram for 2MASS 0030-14, for both Set~1 and Set~2.  The
upper plot is the original Lomb-Scargle periodogram.  The lower  plot is
the same periodogram scaled to confidence units.  Thus a value of 0.8
indicates 80\% confidence that a peak is not due to sampling effects
or random noise.  The dashed line denotes the separation between Set~1 
and Set~2 periodograms. \label{per4}}
\end{figure}

\begin{figure}
\plotone{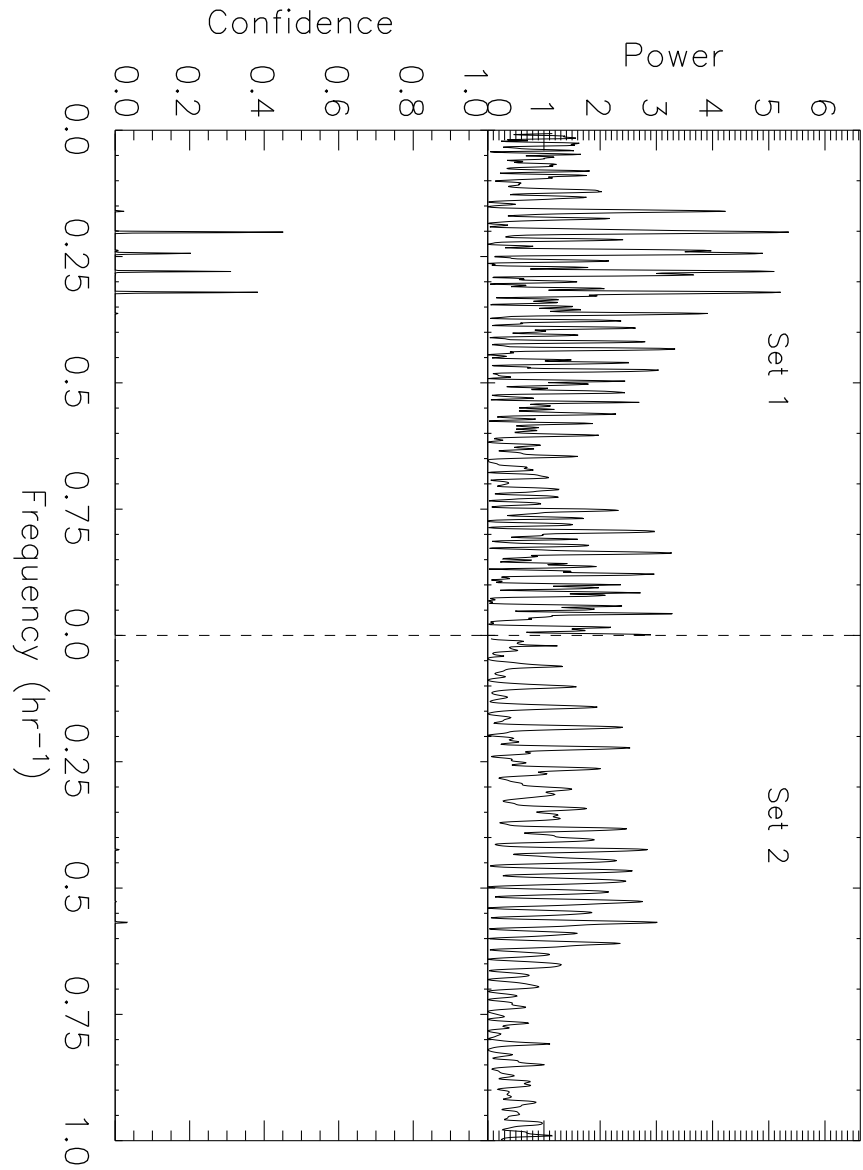}
\caption{Periodogram for the field star of 2MASS 0030-14.  
See Figure~\ref{per4} for explanation. \label{per4_ref}}
\end{figure}

\begin{figure}
\plotone{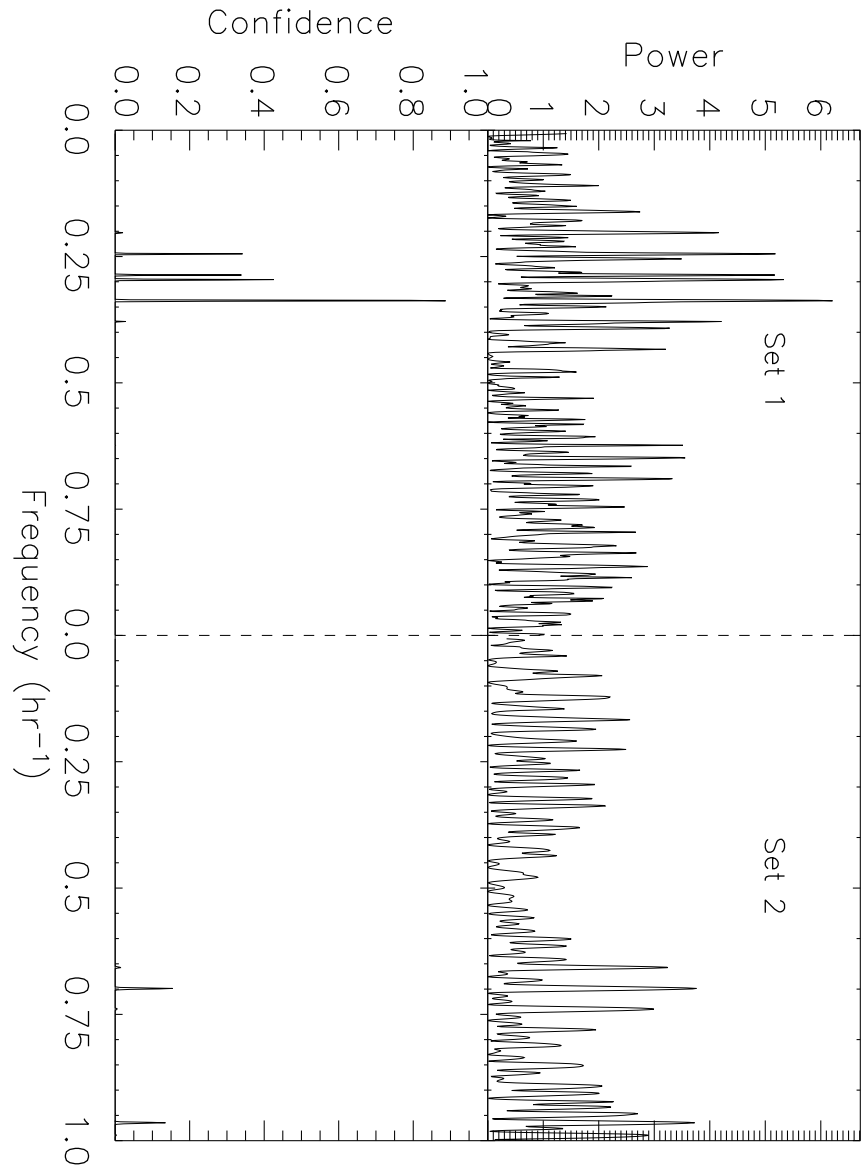}
\caption{Periodogram for SDSS 0151+12.  
See Figure~\ref{per4} for explanation. \label{per5}}
\end{figure}

\begin{figure}
\plotone{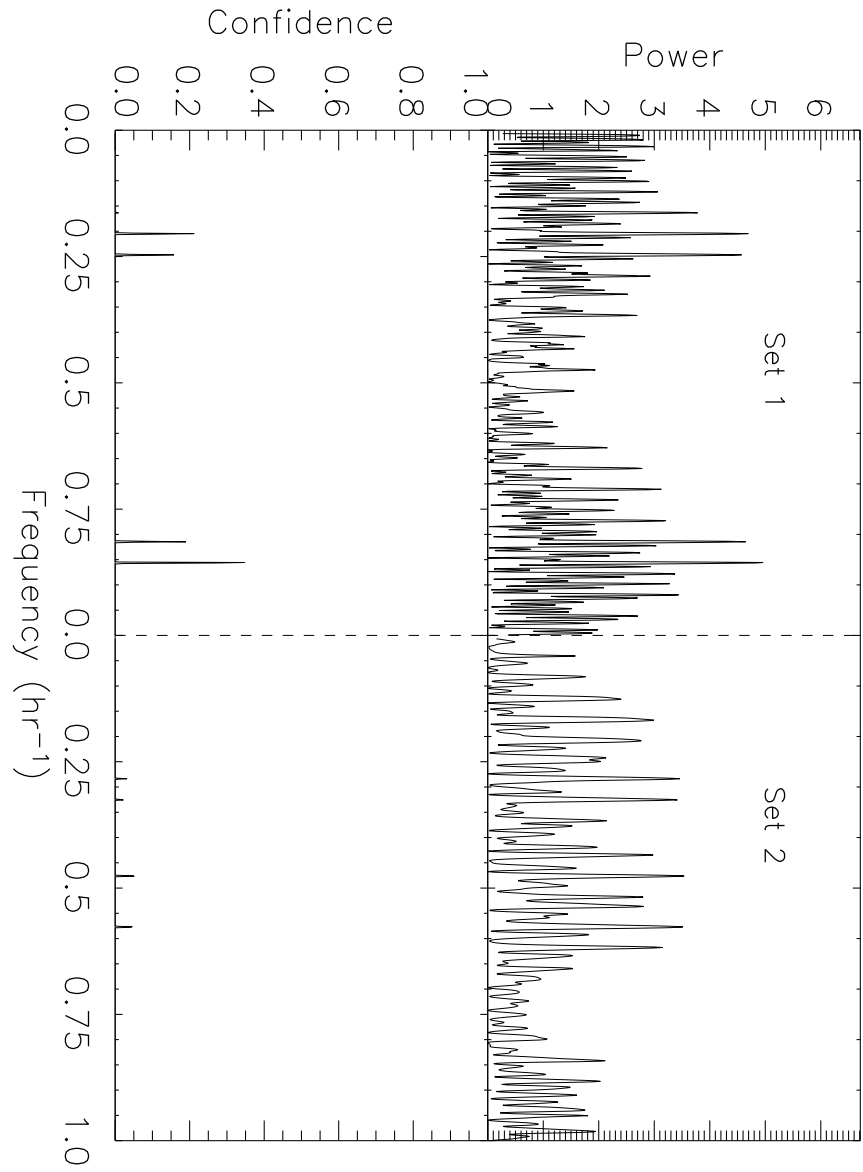}
\caption{Periodogram for the field star of SDSS 0151+12.  
See Figure~\ref{per4} for explanation. \label{per5_ref}}
\end{figure}

Table~\ref{tblper} gives the results of the periodogram analysis for
variable and possibly variable targets.  The best period,
corresponding to the highest peak  in the periodogram, is given in
hours followed by the confidence that the peak is not due to noise or 
sampling effects, as determined by shuffling.  There are no detected periods with 
a confidence $>95\%$ for any of the objects; in fact  only two of the targets,
2MASS~0030-14 and SDSS~0151+12, have periodogram peaks above the  68\%
($\approx1\sigma$) level.   Figures~\ref{per4} and \ref{per5} show the
periodograms of these two targets.  Upper plots
show the original periodograms; in the lower plots each point has been 
replaced by its corresponding  confidence level.  Figures~\ref{per4_ref} 
and \ref{per5_ref} show similar plots for the field stars.   
Were the periodic behavior observed intrinsic to the comparison star rather 
than the target, the field star should also exhibit periodic behavior.  
Thus a lack of strong peaks in the field star periodogram indicates that the 
tentative periodicity is intrinsic to the target.

Another way to distinguish real peaks from spurious ones, apart from
our method  of shuffling,  is to deconvolve the  sampling window
function from the periodogram.  The Fourier transform of the  true
light curve is convolved with the sampling window function during 
observation, causing
false peaks and structure in the periodogram.  The analytic window
function in this case is the Fourier transform of a sum of delta
functions located at the sampling times.  When the data are
deconvolved using a   maximum likelihood routine, it is found that any
significant  peaks present in the deconvolved periodogram almost
always correspond to  the strongest peak in the original periodogram.
For this reason we choose to quantify  the periodogram peaks in terms
of their confidence level (as in Figures~\ref{per4}  and~\ref{per5}), 
which gives us a better basis for  comparing results.

\begin{figure}
\plotone{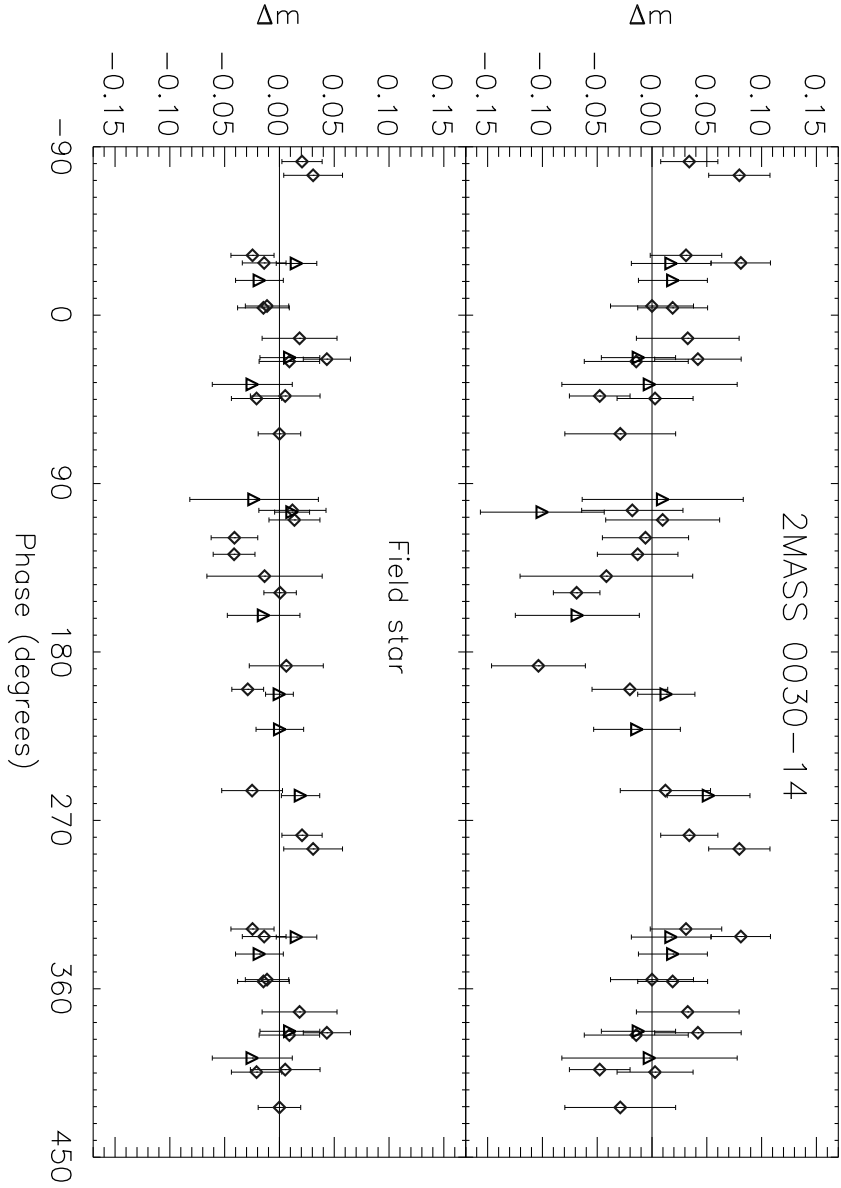}
\caption{Light curve of 2MASS 0030-14, phased to a period of 1.46
hours.  In total, 1.5 periods are plotted.  Diamond points indicate
data from Set~1, and triangle points data from Set~2.  The field star 
light curve is also phased to 1.46 hours for comparison.  \label{phase4}}
\end{figure}

\begin{figure}
\plotone{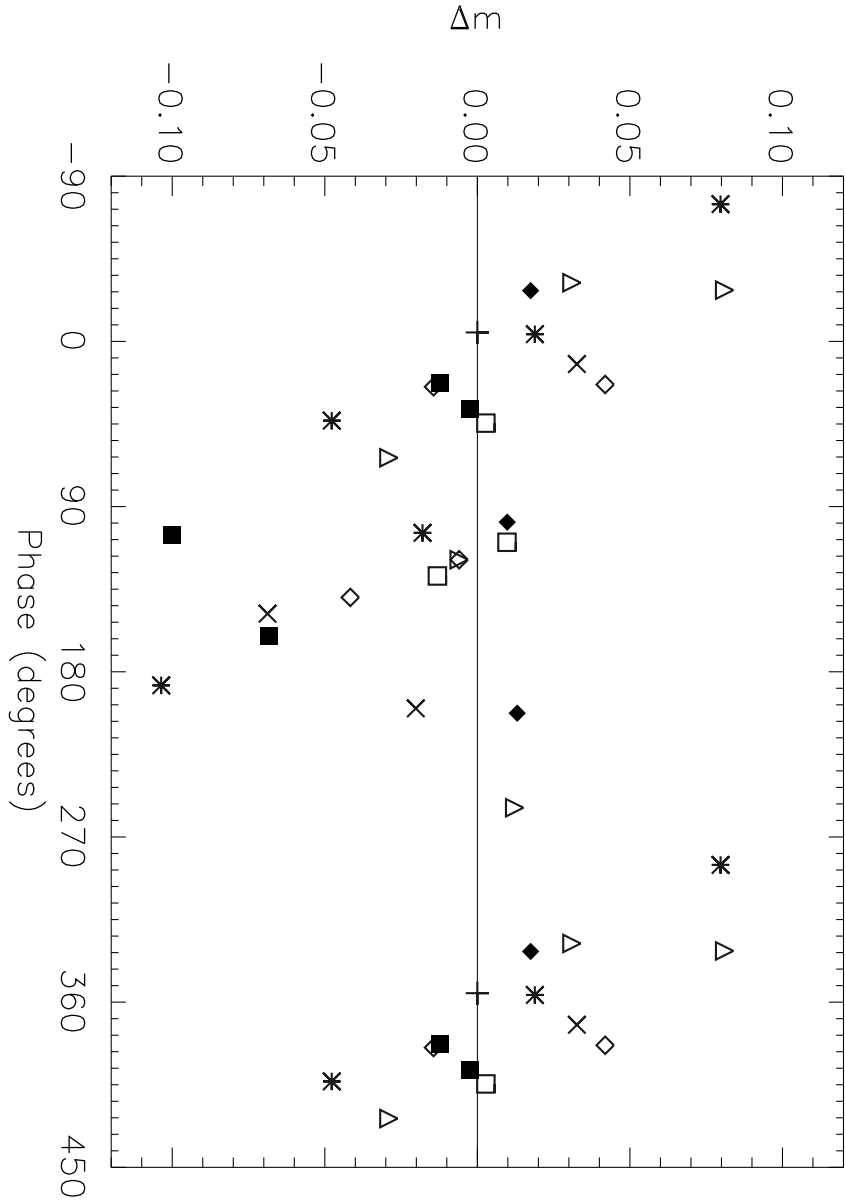}
\caption{Phased light curve of 2MASS 0030-14, similar to 
Figure~\ref{phase4} but here each day of observation is denoted by
a different symbol.  The filled symbols correspond to days in Set~2.
\label{phase4day}}
\end{figure}

\begin{figure}
\plotone{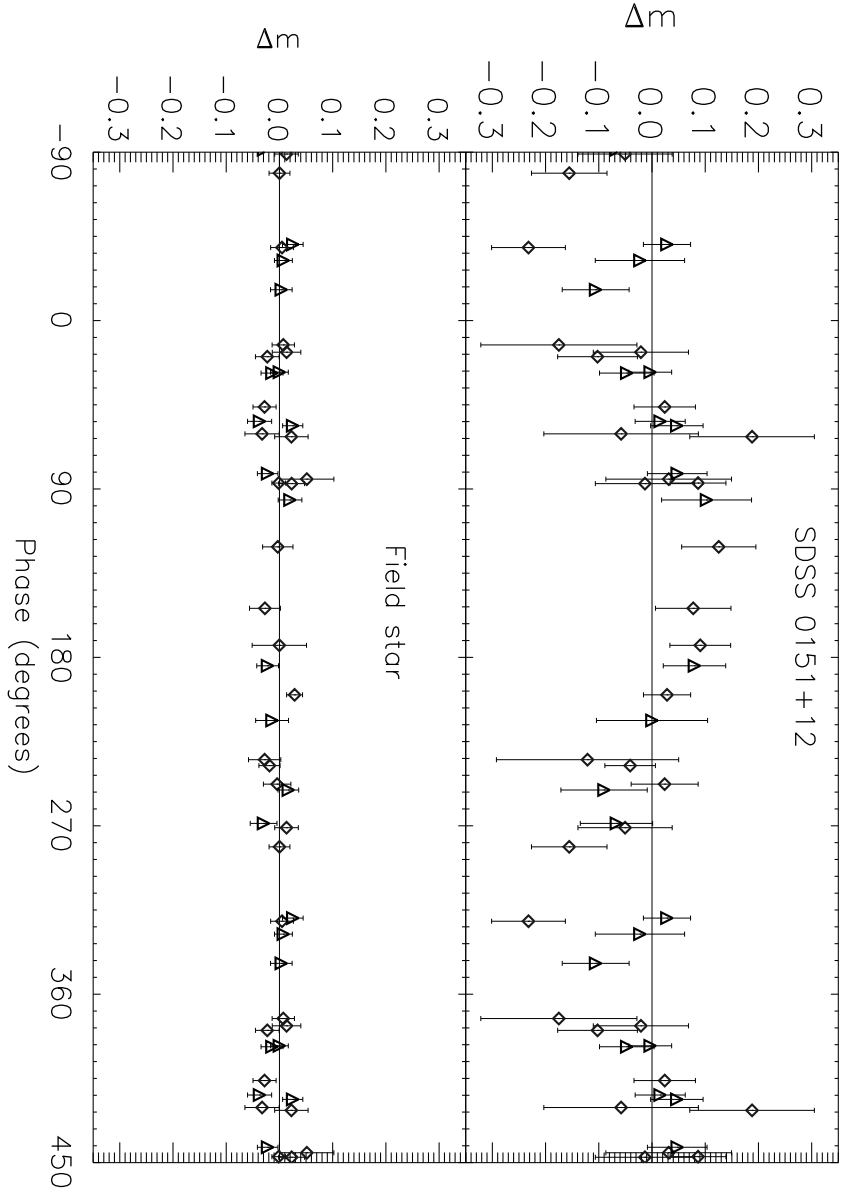}
\caption{Light curve for SDSS 0151+12 phased to a period of 2.97
hours.  See Figure~\ref{phase4} for explanation.  Here the data for
Set~2 have been shifted by 110 degrees of phase to visually align them 
with the Set~1 data.  Note the rise in Set~2
around 330 degrees, which may indicate that either there is an extra
bump in the Set~2 light curve, or the period for Set~2 is not the same 
as that for Set~1.
\label{phase5}}
\end{figure}

\begin{figure}
\plotone{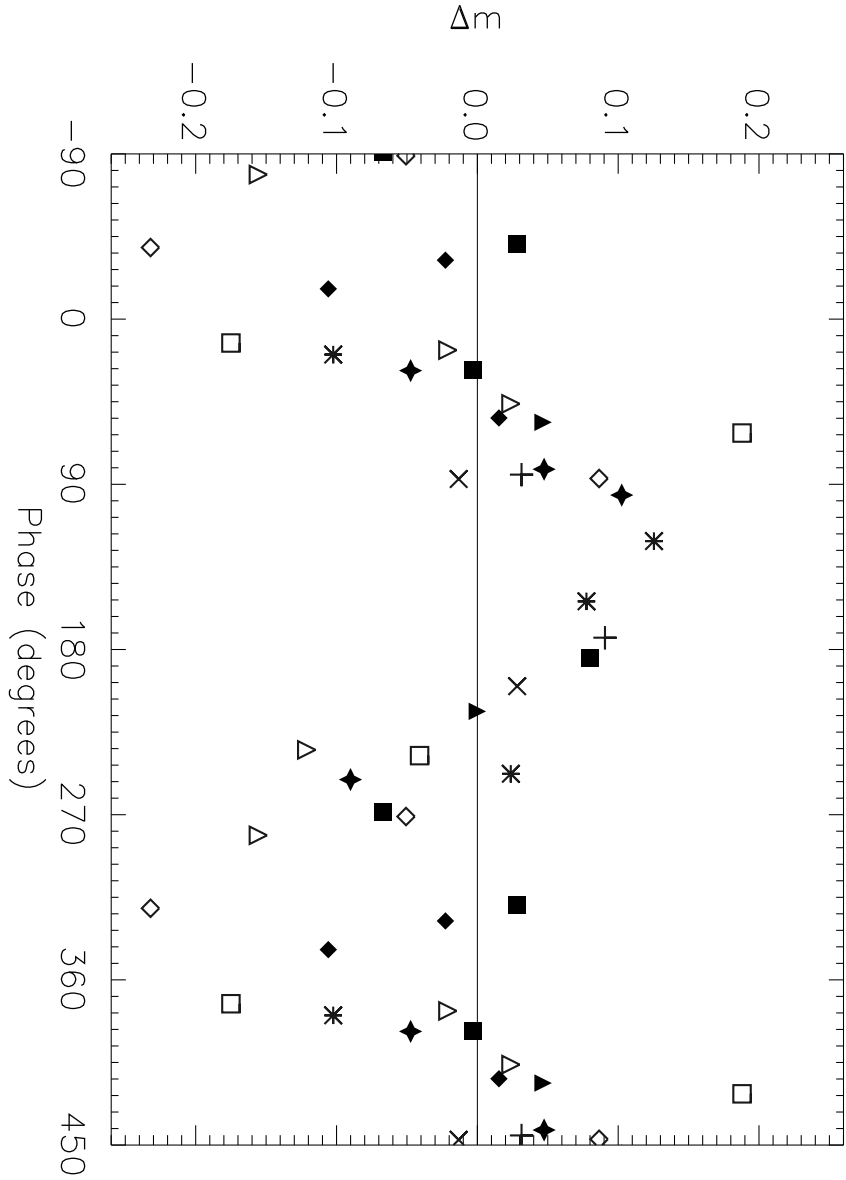}
\caption{Phased light curve of SDSS 0151+12, plotted by day.  See 
Figure~\ref{phase4day} for explanation. \label{phase5day}}
\end{figure}

The light curves of the two periodic detections are phased to their
best period as  determined by the periodogram and plotted in the upper
panels of Figures~\ref{phase4} and \ref{phase5}.  Note that although 
in both cases the best period is determined from the Set~1 data only, 
all the data have been phased to this period in the figures.   In the lower 
panels are the field star light curves, phased to the 
same period as the target (\textit{not} their own best period).
As a further check on the believability of these periods, 
Figures~\ref{phase4day} and \ref{phase5day} 
show each day of observation denoted by a different symbol.

The other objects in Table~\ref{tblper} were also phased to their best
period from the  periodogram and checked by eye, but none displayed any
obviously periodic behavior.  As \cite{bjm01} note, the light curves of 
objects that are variable due to patchy clouds  will not always look
sinusoidal.  Furthermore, it is much more difficult to detect periodic
behavior than it is to detect variability.  Given that none of the
variability detections  have extremely high confidence, we cannot rule 
out the presence of periodic variations in the other variable, or 
possibly variable targets.

\subsubsection{Discussion of individual objects: Possible detections}\label{indper}

\textbf{2MASS 0030-14:}
There are two close peaks (at 1.38 and 1.46 hours, or 0.72 and 0.68 hr$^{-1}$) 
in the 2MASS 0030-14 periodogram at approximately the same (90\%) confidence 
level (Figure~\ref{per4}).
Either of these could be considered the best period; we have chosen to use the 
1.46 hour period. The next highest peak, at 6.67 hours (0.15 hr$^{-1}$) and 
85\% confidence, also shows possible periodic behavior, but not  as clearly as 
for 1.46 hr.  Note that the field star periodogram (Figure~\ref{per4_ref}) 
never exceeds 50\% confidence.
The light curve of 2MASS 0030-14, phased to 1.46 hours (Figure~\ref{phase4}), 
is not exactly sinusoidal, but the periodic nature is evident.   The
amplitude of the curve indicates a fractional deviation from the mean
flux of $\sim10\%$.  
The periodic behavior is indeed intrinsic to the
brown dwarf and not to the comparison  star, as can be seen by
examining the field star light curve phased to the same period.   
Also, note that in Figure~\ref{phase4day} the data from each day follow 
approximately the same curve without any systematic trends,
supporting the conclusion that the observed periodicity is real.
The data from Set~1 (diamond points) and Set~2 (triangle points) match 
well without any phase shift between them; if this is the 
correct period it is probably stable over the length of our observations.  

\textbf{SDSS 0151+12:}    The periodogram for SDSS 0151+12 
(Figure~\ref{per5}) has only one significant peak, at a period of 
2.96 hours (0.34 hr$^{-1}$) and a confidence of 89\%.  The phased light 
curve (Figure~\ref{phase5}) shows a nearly sinusoidal curve for the 
Set~1 data.  The Set~2 data, however, have an extra peak around 330 degrees.   
This peak is present in at least two of the nights in Set~2, as is evident 
from Figure~\ref{phase5day}.  We propose three possible reasons for the
slight mis-match  between Set~1 and Set~2: the light curve changed 
between the two sets with an increase in flux appearing
around 330 degrees, the period of variation changed  between
the two sets so that the phased light curve no longer shows one
period, or the derived period of variation is incorrect.  
It is clear that a change in the light curve could cause the Set~2 data  
to have a smaller overall amplitude and help explain why SDSS 0151+12 
is not a  detection in Set~2.  
The amplitude of the Set~1 data corresponds to a fractional deviation from
the mean flux of  $\sim20\%$.
Note that the Set~2 data have been shifted by 110
degrees of phase in order to visually align them with the Set~1 data.
Such a phase offset is not unexpected; even a small 
error in the period can cause large phase offsets because the 
observational baseline is long compared to one period.
For example, if we make a 1\% error in a 3 hour period, the 420 hour 
time span between Set~1 and Set~2 will introduce a phase offset of 
500 degrees.

\section{Discussion}\label{discussion}

Our result, that at least one third of our targets are variable 
(and as many as two thirds if possible detections are accounted for), is 
consistent with previous surveys of L dwarf variability.  
\cite{bjm01} find 7 of their 10 early L dwarfs to be variable, 
\cite{gel02} find 7 of 18 variable, and \cite{clark02b} find 2 of 4
variable.  However, \cite{bjl02} fail to find variability at the 
0.04 mags level in the $J$ and $K'$ bands for three early L dwarfs.
Given these statistics and our current results, there is 
no evidence that objects near the L/T boundary
are more likely to be variable than  earlier L dwarfs.  It is 
clear, however, that variability does exist in even these very cool objects.  

As most of the previous studies were conducted in the $I$-band, it is
impossible to quantitatively compare detection limits with our $K_s$ band
observations.   We note, however, that the detection limits of \cite{bjl02} 
are well below our observed amplitudes.  The fractional amplitudes of variation 
in the $I$-band for most early L dwarfs is small, with  few above the 1-2\%
level~\citep{clark02b}.  Our variable objects show much larger 
deviations, but again such comparison across wavebands are difficult to 
quantify.  Further studies at infrared wavelengths will need to be done 
before we can meaningfully compare either the fraction of variable objects 
or the variability amplitudes of early L dwarfs and L/T transition objects.

In light of the cloud-clearing model fit \citep{burg02a}, we might 
expect the photometric signal to be rotationally modulated.  Therefore, 
if the signal were due to a single atmospheric feature, the period of 
variation would be equal to the rotational period.
Were the 1.5 hour variation of 2MASS 0030-14 a true measurement of the 
rotational period, the rotational velocity of this object would lie at 
the high end of observations \citep{bas00}, but it is well below breakup 
velocity for any M$>0.01$M$_{\sun}$.  The 3.0 hour variation of SDSS 0151+12 
would also translate to fairly rapid rotation. 
In the context of this picture we might also expect a trend in variability 
with spectral type, although the relationship would not be linear.  If cloud break-up 
occurs at the transition from L to T dwarfs, the variability should peak at 
that point, falling off toward both mid-L and mid-T dwarfs.  

\begin{figure}
\plotone{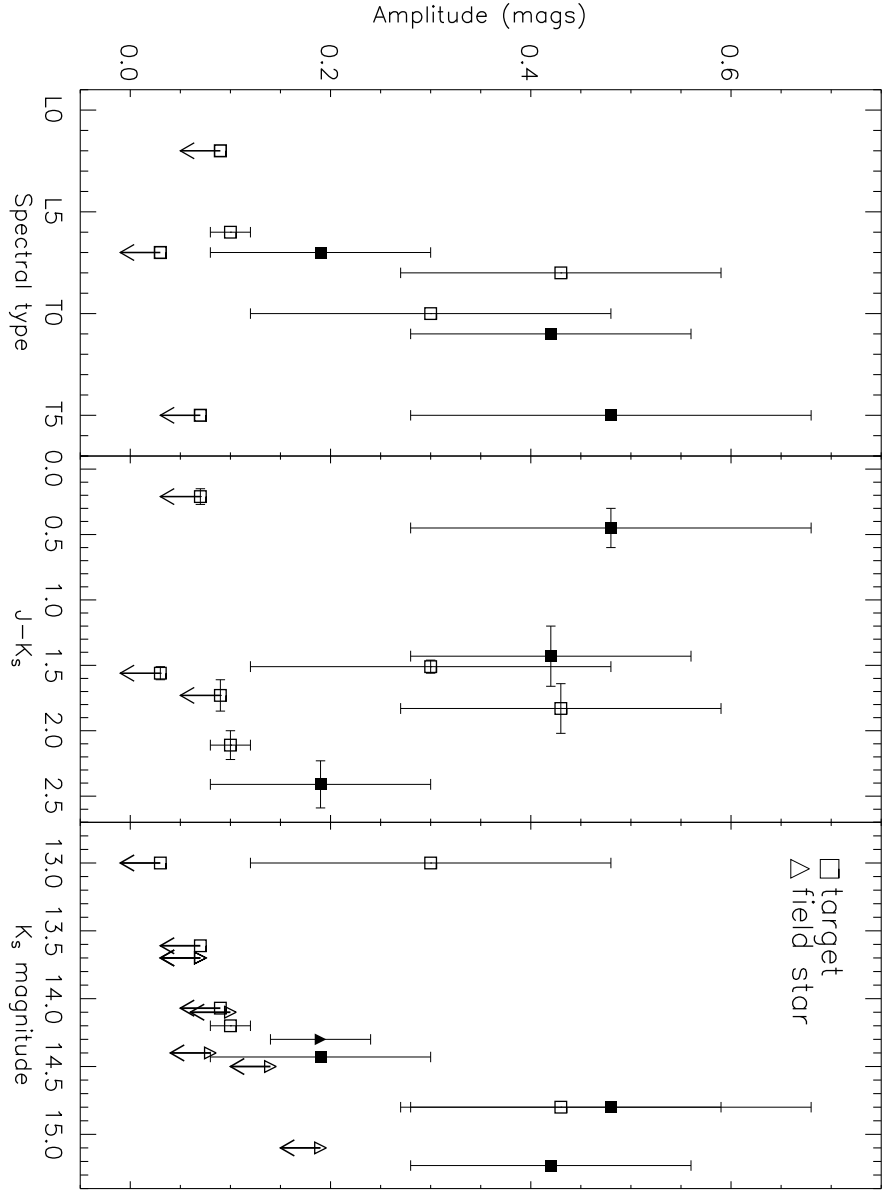}
\caption{Variability amplitudes and detection limits from
Table~\ref{tblamps} vs spectral type, $J-K_s$ color, and $K_s$ 
magnitude.  Left: squares with error bars 
indicate variability amplitudes for detections (solid squares) and 
possible detections (open squares).  For non-detections the 99\% 
sinusoidal detection limits are denoted by  open squares with arrows.  
Center:  $J-K_s$ values and errors are from references in Table~\ref{tbl1}. 
Right: similar to left panel, but here field star 
amplitudes and detection limits are included as triangle symbols.  
The $K_s$ magnitude of a field star is computed from the 
average differential photometry: 
$(K_s)_{\mathrm{field}} =(K_s)_{\mathrm{target}}-\overline{\Delta \mathrm{m}_{\mathrm{target}}} + \overline{\Delta \mathrm{m}_{\mathrm{field}}}$. \label{correl}}
\end{figure}

In Figure~\ref{correl} we plot amplitudes of variability and
detection limits as a function of spectral type (left), $J-K_s$ color (center), 
and $K_s$ magnitude (right).  
There is no clear visual evidence to suggest that L/T boundary objects are 
more likely to be variable than later or earlier type objects.
To test for a statistical correlation between spectral type and variability amplitude 
we compute the Spearman rank correlation coefficient ($-1 \le \rho_s \le 1$) 
and significance value ($0\le \zeta \le 1$).  A value of $\rho_s=1$ implies 
perfect correlation, and the smaller $\zeta$ the more significant the 
result.  Note that for the following, non-detections are assumed to have variability 
amplitudes of zero.\footnote{ Another possibility is to let non-detection amplitudes 
equal their detection limits.  We find that the two methods give very similar results.} 
For amplitude vs spectral type we find 
$\rho_s=0.48$, which corresponds to a significance of $\zeta=0.19$ and a 
two-sided null hypothesis probability ($P_{null}$) of about 18\%.  Thus there is no strong 
evidence for a correlation between variability amplitude and spectral type.

When we apply the rank correlation test to variability amplitude vs $J-K_s$ 
color we find no correlation ($P_{null}=80\%$), as verified by an examination of 
Figure~\ref{correl}.
A similar comparison of amplitude vs $K_s$ mag, however, yields values of 
$\rho_s=0.71$, $\zeta=0.03$, and $P_{null}=4\%$, thus the two properties are 
correlated with $\sim96\%$ confidence.   It is unclear why magnitude and
variability would be correlated, given that apparent
magnitude is primarily a function of distance, not physical
properties.  While this may seem to indicate a problem with our analysis, 
we note that no such trend is seen for the field stars (see Figure~\ref{correl}), 
which have comparable $K_s$ magnitudes.  This discrepancy suggests that there 
may be a hidden dependency causing these correlations.  If, for example, all the 
targets were at the same distance, we would expect apparent magnitude and 
spectral type to be well correlated.  
Curiously, if we neglect the three
brightest objects there are strong correlations ($P_{null}\le 7\%$) between
amplitude and spectral type, amplitude and $K_s$ magnitude, \textit{and}
magnitude and spectral type.  
We suspect that complex selection effects involving target selection, 
apparent magnitude, and spectral type are complicating these results.
A larger sample is
needed to disentangle these dependencies and determine if there are any 
trends of variability frequency or amplitude with spectral type.

\begin{figure}
\plotone{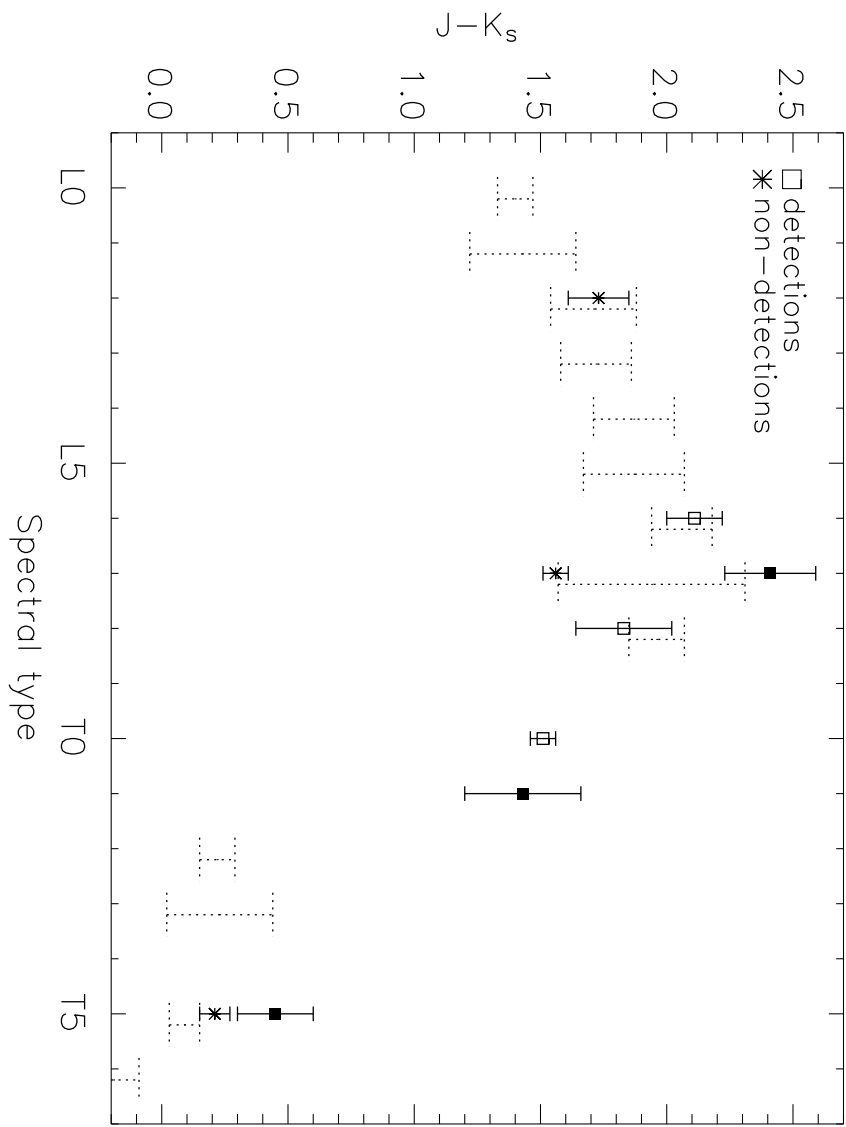}
\caption{ The $J-K_s$ colors of targets as a function of spectral type.  
Solid squares indicate detections, open squares possible detections, and 
stars non-detections.  Dotted bars are average 2MASS $J-K_s$ values, with 
errors, from \protect\cite{kirk00} and \protect\cite{burg02b}.  \label{jkplot}}
\end{figure}

As \cite{gel01} point out,  if holes in
the cloud deck are responsible for photometric  variability in brown
dwarfs, we might expect the $J-K_s$ colors of variable objects to be
bluer than those of non-variable objects (which are assumed to
have uniform cloud coverage) for a given spectral type.   
In Figure~\ref{jkplot} we plot $J-K_s$ vs spectral type for our targets, 
as well as the average $J-K_s$ as a function of spectral type from 
\cite{kirk00} and \cite{burg02b}.
Unfortunately, there are so few $J-K_s$ measurements for L9-T2 objects
that there are no average colors available.  
From this plot there is no indication that variable objects
have bluer $J-K_s$ colors compared to non-variable objects.  
There is also no evidence that variable objects are bluer than the 
average $J-K_s$ for their spectral type.
If, however, at least some non-variable brown
dwarfs do not vary  because the holes or structure in their atmosphere
is more uniform (e.g. many small holes),  this correlation would not
be so simple.  

It has recently been suggested that any observed IR variability of brown dwarfs
may be due entirely to second order telluric extinction effects \citep{bjl02}.  
The unique spectral energy distribution of brown dwarfs coupled with 
wavelength-dependent atmospheric extinction may cause
different fractional decreases in intensity with airmass across a broad band
filter.  This effect could cause the magnitude difference between brown dwarf 
and comparison star to change over the course of observation, creating an artificial 
variability signal.  
To check for such a bias we examined
plots of $\Delta$m vs airmass for each object.  Again applying the rank correlation
test, we find that for most objects $P_{null}\ge 60\%$.
The lowest null probability, 
for SDSS~0151+12, is 22\%, but when the highest airmass point is removed (which does
not affect any other results) $P_{null}$ rises to $>40\%$.  Therefore, we find no evidence
that $\Delta$m is correlated with airmass for any of the targets, and argue that 
changing airmass is not responsible for the observed variability.
Furthermore, it seems unlikely that atmospheric effects could produce a
periodic signal as is seen here in two objects.  
Although we recognize that extinction effects could possibly complicate some of our 
detections, we remain confident of our general results.

\section{Summary}\label{sum}

We have obtained $K_s$-band light curves for nine late L and early T
dwarfs.  Results  indicate that three show significant
evidence for variability, and several more show evidence for possible 
variability, with approximate  fractional deviations from the mean 
flux of 5\% to 25\% (peak-to-peak amplitudes of 0.1 to  0.5 mags).  
Two objects, 2MASS 0030-14 (L7) and SDSS 0151+12 (T1), have marginally
significant peaks in the Lomb-Scargle periodogram.  To these objects 
we assign tentative periods of 1.5 and 3.0 hours respectively.  
In the case of SDSS 0151+12, it seems likely that either the period of 
variation or the shape of the light curve changed over the 29 day 
observational baseline.

While it is clear that variability exists in these cool objects, the 
source of variability is still unknown.  As noted in Section~\ref{intro}, 
it seems unlikely that magnetic activity could account for the variability 
seen here.  We see no significant trend of variability with spectral type, which 
might be expected were the break-up of cloud decks responsible.  Similarly, 
there is no evidence that variable objects are bluer than non-variable 
objects of the same spectral type.
Time resolved spectroscopic measurements of variable targets might 
be able to conclusively determine whether the variability seen in this 
survey is due to structure in the cloud coverage or some other phenomenon.

\acknowledgments

The authors are grateful to Antonin Bouchez, Jean-Luc Margot, and the Palomar 
Observatory staff for their time and help in obtaining these observations.  
We also thank the referee for helpful comments on the manuscript.
M. L. E. acknowledges support from a NSF Graduate Research Fellowship.
A. J. B.  acknowledges support by NASA
through Hubble Fellowship grant HST-HF-01137.01 awarded by the
Space Telescope Science Institute, which is operated by the
Association of Universities for Research in Astronomy, Inc., for
NASA, under contract NAS 5-26555.
This research has made use of
the NASA/IPAC Infrared Science Archive, which is operated by the
Jet Propulsion Laboratory, California Institute of Technology,
under contract with the National Aeronautics and Space
Administration. This publication makes use of data from the Two
Micron All Sky Survey, which is a joint project of the University
of Massachusetts and the Infrared Processing and Analysis Center,
funded by the National Aeronautics and Space Administration and
the National Science Foundation.

\clearpage

\epsscale{0.85}

\epsscale{0.75}

\clearpage

\begin{deluxetable}{llccc}
\tabletypesize{\small} 
\tablecaption{Properties of L and T dwarf targets  \label{tbl1}} 
\tablewidth{0pt} 
\tablehead{  \colhead{Target name}    &  \colhead{IAU name}    &  
\colhead{Type} &  \colhead{$K_s$ (2MASS)}  &  \colhead{Reference}   } 

\startdata

2MASS 2208+29  & 2MASSW J2208136+292121  & L2p & 14.09$\pm$0.08 & 1  \\  
2MASS 0103+19  & 2MASSI J0103320+193536  & L6  & 14.15$\pm$0.07 & 1  \\  
DENIS 0205-11  & DENIS-P J0205.4-1159AB  & L7  & $13.0\pm0.2$\tablenotemark{a} & 2 \\ 
2MASS 0030-14  & 2MASSW J0030300-145033  & L7  & 14.38$\pm$0.08 & 1 \\  
2MASS 0328+23  & 2MASSI J0328426+230205  & L8  & 14.84$\pm$0.13 & 1 \\  
SDSS 0423-04   & SDSSp J042348.57-041403.5 & T0\tablenotemark{b}  & 12.94$\pm$0.04 & 3 \\  
SDSS 0151+12   & SDSSp J015141.69+124429.6   & T1\tablenotemark{b}  & 15.09$\pm$0.19 & 3 \\  
2MASS 2254+31  & 2MASSI J2254188+312349  & T5  & 14.83$\pm$0.14 & 4  \\  
2MASS 0559-14  & 2MASSI J0559191-140448  & T5  & $13.61\pm0.05$ & 5 \\

\enddata 
\tablecomments{Spectral types from \cite{kirk99, kirk00} and \cite{burg02b} 
unless otherwise noted.  $K_s$ magnitudes are from 2MASS photometry.}
\tablenotetext{a}{Magnitude is for the binary pair.}  
\tablenotetext{b}{Spectral type assigned by \cite{geb02}.}
\tablerefs{(1)~\cite{kirk00}; (2)~\cite{del97}; (3)~\cite{geb02}; (4)~\cite{burg02b}; (5)~\cite{burg00} }

\end{deluxetable}

\clearpage

\begin{deluxetable}{lccccccccccr}
\tablecaption{Results of variability analysis \label{tblvar}}
\tablewidth{0pt}
\tabletypesize{\scriptsize} 
\tablehead{ \colhead{Target } &  \colhead{Spectral}   & & 
\multicolumn{2}{c}{ Set~1 } & & \multicolumn{2}{c}{ Set~2 } & & 
\multicolumn{2}{c}{ Combined set }  \\ 
\cline{4-5} \cline{7-8} \cline{10-11} 
\colhead{ name } &  \colhead{type}  & &  \colhead{$\widetilde{\eta}$} & 
\colhead{conf}  & & \colhead{$\widetilde{\eta}$} &  \colhead{conf}  &  &
\colhead{$\widetilde{\eta}$ } & \colhead{conf} } 

\startdata 
2MASS 2208+29 & L2p & &  0.76 & 34\% & & 0.82 & 54\% & & 0.79 & 44\% &  \\
2MASS 0103+19 & L6  & &  0.88 & 69\% & & 1.48 & $>99\%$\tablenotemark{a} & & 1.06 & 98\%\tablenotemark{a} & \\  
DENIS 0205-11 & L7 & & 0.68  &  18\% & &  1.00 & 87\%  & & 0.80  & 48\% & \\  
\textbf{2MASS 0030-14} & \textbf{L7} & & \textbf{1.09} & \textbf{98\%} & & 
0.79 & 33\%  & & \textbf{1.00}  & \textbf{95\%} &  \\  
2MASS 0328+23 & L8 & & 1.43 & $>99\%$\tablenotemark{b} & & 
1.03  & 88\%  & & 1.21 & $>99\%$\tablenotemark{b} &  \\  
SDSS 0423-04  & T0 & & 0.83  &  56\%  & & 1.04  & 87\% & & 1.01 & 96\%\tablenotemark{c} &  \\  
\textbf{SDSS 0151+12} & \textbf{T1}  & & \textbf{1.15} & \textbf{99\%}  & &  
0.87 & 64\%  & &  \textbf{0.99}  & \textbf{95\%}    &   \\  
\textbf{2MASS 2254+31} & \textbf{T5} & &  0.92   & 84\%  & & 1.00  & 88\%  & &
\textbf{0.99} & \textbf{98\%} &  \\   
2MASS 0559-14 & T5  & & 0.78  &  45\% & & 0.84  & 55\%  & & 0.80 & 47\% &  \\

\tableline \multicolumn{12}{c}{Field stars}   \\ 
\tableline \\

2MASS 2208+29  &  & &  0.84  & 62\%  & &  0.86  & 62\% & & 0.84  & 66\%  & \\  
2MASS 0103+19  &  & &  0.83  &  57\%  & &  0.95 & 75\% & & 0.84  & 67\%  & \\  
DENIS 0205-11  &  & &  \nodata  &  \nodata  & &  \nodata  & \nodata   & & 
\nodata & \nodata & \\  
2MASS 0030-14  & & &  0.85 &  62\%  & &  0.58  & 14\% & & 0.78  & 44\%  & \\  
2MASS 0328+23  &  & &  \nodata  & \nodata   & &  \nodata  & \nodata   & &
\nodata & \nodata & \\  
\textbf{SDSS 0423-04} &  & & \textbf{1.10} & \textbf{97\%}& & \textbf{1.33} & 
$\mathbf{99\%}$ & & \textbf{1.18} & $\mathbf{>99\%}$ & \\  
SDSS 0151+12   &   & &  0.71  & 25\%  & & 0.93 & 75\%  & & 0.77 & 37\%  & \\  
2MASS 2254+31  &   & &  0.81  & 51\%  & & 0.97 & 82\%  & & 0.82 & 59\%  & \\  
2MASS 0559-14  &   & &  0.73  & 32\%  & & 0.97 & 76\%  & & 0.80  & 52\%  & \\

\enddata


%

\tablecomments{$\widetilde{\eta}$ is the reduced robust median
statistic, and conf is the percentage of times that an intrinsically
non-variable source with random errors had a smaller
$\widetilde{\eta}$.  
We take a confidence $>95$\% to indicate a
variability detection; such detections are shown in bold-face type.
Possible detections (see notes and Section~\ref{possvar}) are left in plain type.
For two of the targets no field star was available.}

\tablenotetext{a}{When the largest outlier is removed these confidences 
drop below the 95\% cutoff.}
\tablenotetext{b}{No field star was available for 2MASS 0328+23, thus we 
are unable to confirm that variations are not due to the comparison star.}
\tablenotetext{c}{Variability could be due to a variable comparison star.  
Note the variability detection in the field star section.}

\end{deluxetable}

\clearpage

\begin{deluxetable}{lcccccccccr}
\tabletypesize{\scriptsize} 
\tablecaption{Variability amplitudes and 99\% detection limits \label{tblamps}} 
\tablewidth{0pt} 
\tablehead{ \colhead{Target } & & \multicolumn{2}{c}{ Set~1 } & &
\multicolumn{2}{c}{ Set~2 } & &
\multicolumn{2}{c}{ Combined set } & \\ 
\cline{3-4} \cline{6-7} \cline{9-10} 
\colhead{ name }           & &  \colhead{amp} &  
\colhead{det limit}  & &
\colhead{amp} &  \colhead{det limit}  &  & \colhead{amp} &
\colhead{det limit} &  \\ 
\colhead{ \ }           & &  \colhead{(mags)} &  
\colhead{(mags)}  & &
\colhead{(mags)} &  \colhead{(mags)}  &  & \colhead{(mags)} &
\colhead{(mags)} &
} 

\startdata 

2MASS 2208+29 & & \nodata & 0.10 (0.15) & & \nodata & 0.17 (0.16) & & 
\nodata & 0.09 (0.13) & \\  
2MASS 0103+19  & & \nodata & 0.09 (0.13) & & $0.10\pm0.02$ & 
0.13 (0.12) & & $0.10\pm0.02$ & 0.08 (0.11) &  \\
DENIS 0205-11& & \nodata & 0.04 (0.05) & & \nodata & 0.04 (0.06) & & 
\nodata & 0.03 (0.05) &  \\  
\textbf{2MASS 0030-14}  & & $0.19\pm0.11$ & 0.14 (0.19) & & \nodata & 
0.22 (0.22) & & $0.19\pm0.11$ & 0.12 (0.20) &  \\  
2MASS 0328+23  & & $0.32\pm0.07$ & 0.27 (0.32) & & \nodata & 
0.23 (0.33) & & $0.43\pm0.16$ & 0.21 (0.29) &  \\  
SDSS 0423-04 & & \nodata & 0.14 (0.18) & & \nodata & 0.08 (0.10) & & 
$0.30\pm0.18$ & 0.08 (0.11) &  \\  
\textbf{SDSS 0151+12}   & & $0.42\pm0.14$ & 0.25 (0.41) & & \nodata & 
0.30 (0.33) & & $0.42\pm0.14$ & 0.22 (0.35) & \\  
\textbf{2MASS 2254+31}  & & \nodata & 0.21 (0.33) & & \nodata  & 
0.32 (0.30) & & $0.48\pm0.20$ & 0.20 (0.28) &  \\   
2MASS 0559-14 & & \nodata & 0.09 (0.11) & & \nodata & 0.20 (0.20) & & 
\nodata & 0.07 (0.11) &  \\

\tableline \multicolumn{11}{c}{Field stars}   \\ 
\tableline \\

2MASS 2208+29   & & \nodata & 0.14 (0.22) & & \nodata & 0.25 (0.24) &
& \nodata & 0.14 (0.19) & \\  
2MASS 0103+19   & & \nodata & 0.20 (0.31) & & \nodata & 0.32 (0.30) & & 
\nodata & 0.19 (0.28) & \\  
DENIS 0205-11   & & \nodata & \nodata & & \nodata & \nodata & & \nodata &
\nodata & \\
2MASS 0030-14  & & \nodata & 0.08 (0.13) & & \nodata & 0.09 (0.13) & & 
\nodata & 0.07 (0.11) & \\  
2MASS 0328+23  & &  \nodata &  \nodata & &  \nodata &  \nodata & &  \nodata & 
\nodata &  \\
\textbf{SDSS 0423-04}   & & $0.17\pm0.06$ & 0.14 (0.18) & & $0.14\pm0.04$ & 
0.14 (0.17) & & $0.19\pm0.05$ & 0.10 (0.16) & \\
SDSS 0151+12    & & \nodata & 0.08 (0.13) & & \nodata & 0.10 (0.13) &
& \nodata & 0.07 (0.12) & \\  
2MASS 2254+31   & & \nodata & 0.09 (0.14) & & \nodata & 0.16 (0.17) & & 
\nodata & 0.08 (0.14) & \\  
2MASS 0559-14   & & \nodata & 0.12 (0.17) & & \nodata & 0.16 (0.20) & &
\nodata & 0.10 (0.15) & \\

\enddata

\tablecomments{Amplitudes are peak-to-peak and given in
magnitudes.  The uncertainty in the two data points used to compute
the amplitude are added in quadrature to estimate the amplitude 
uncertainty.  Detection limits are the peak-to-peak amplitude 
for which 99\% of sinusoidal variations are detected.  
Numbers in parentheses are similar, but for random variations. }

\end{deluxetable}

\clearpage

\begin{deluxetable}{lccccccr}
\tablecaption{Results of Lomb-Scargle
periodogram analysis \label{tblper}} 
\tablewidth{0pt} 
\tablehead{
\colhead{Target } & & \multicolumn{2}{c}{ Set~1 } & & 
\multicolumn{2}{c}{ Set~2 } & \\
\cline{3-4}  \cline{6-7} 
\colhead{ name }           & & \colhead{period (hrs)}    & 
\colhead{confidence}  & & \colhead{period (hrs)} &  \colhead{confidence}  & 
}

\startdata 

2MASS 0103+19 & & 1.95   &  42\%  & & 2.24   &  58\%  & \\
\textbf{2MASS 0030-14}  & & \textbf{1.38 }  &  \textbf{90\%}  & & 
10.7   &  10\% & \\ 
2MASS 0328+23  & & 3.87   &  5\%  & & 2.10   &  $<1\%$  & \\ 
SDSS 0423-04   & & 1.39   &  $<1\%$  & & 1.10   &  11\%  & \\
\textbf{SDSS 0151+12}  & & \textbf{2.97}  &  \textbf{89\%}  & & 
1.43  &  15\% & \\ 
2MASS 2254+31 & & 1.21   &  4\%  & & 2.65   &  9\%  & \\ 

\tableline \multicolumn{8}{c}{Field stars}  \\ 
\tableline \\

2MASS 0103+19  & &    2.14   &  10\%  & &   3.94   &  6\% &  \\ 
2MASS 0030-14  & &   4.96   &  45\%  & &   1.76   &  3\% &  \\ 
2MASS 0328+23  & &   \nodata  & \nodata & & \nodata  &  \nodata & \\
SDSS 0423-04   & & 1.62   &  46\%  & & \textbf{1.00} & \textbf{99\%}  & \\
SDSS 0151+12  & & 1.17 &  34\% & & 2.10 &  5\% & \\  
2MASS 2254+31  & &   1.17   & 9\%  & &  3.87  &  3\%  & \\

\enddata

\tablecomments{Only targets that were found to be variable or possibly variable
(see Table~\ref{tblvar}) are shown.}

\end{deluxetable}






\end{document}